# Concepts of inertial, gravitational and elementary particle masses

S.O. Tagieva[1] and M. Ertürk[2]

[1]Academy of Science, Physics Institute, Baku 370143, Azerbaijan Republic

[2]Department of Physics, Faculty of Arts and Sciences, Onsekiz Mart University, Çanakkale, Turkey

**Abstract**

In this article the concept of mass is analyzed based on the special and general relativity theories and particle (quantum) physics. The mass of a particle ($m = E_0/c^2$) is determined by the minimum (rest) energy, which is necessary for create that particle and which is invariant under Lorentz transformations. The mass of a bound particle in the any field is described by $m < E_0/c^2$ and for free particles in the non-relativistic case the relation $m = E/c^2$ is valid. This relation is not correct in general for particles, and it is wrong to apply it to the fields if we in mainly adopt as basis the mass determination in particles physics. In atoms or nuclei (i.e. if the energies are quantized) the mass of the particles changes discretely. In non-relativistic cases, mass can be considered as a measure of gravitation and inertia similar to the $E/c^2$ in the relativistic case.

## 1. Introduction

Lev Okun published an article about "on the concept of mass in the special relativity" [1] to show that the formula

$$E_0 = mc^2 \tag{1}$$

is correct but the formulas

$$E = mc^2, \tag{2}$$

$$E_0 = m_0 c^2, \tag{3}$$

$$E = m_0 c^2 \tag{4}$$

and

$$m = \frac{m_0}{\sqrt{1 - v^2/c^2}} \tag{5}$$

are wrong. In fact, Eqs. (2-5) are historical artefacts but, at present, they are seen in many textbooks. Mass is invariant under Lorentz coordinate transformations. In reality, there are a lot of physicists who take a position against Okun and the other particle physicists. For example, among the physicists working on general relativity Khrapko thinks differently than Okun and other particle physicists [2]. However, we can not directly measure masses of particles under the small and large velocities and comprise them. Under the small velocities, masses of same elementary particles estimates from the measured ratio $e/m$, which is give as masses of particles, because the value of electric charge are known. But under the relativistic cases we may measure only impulse and energies of particles, and then estimated of masses from relativistic dependents between this quantities and masses. Therefore, the second method may give mass which is determined as $E/c^2$ and also is invariant under the Lorentz transformations or rest mass $E_0/c^2$.

The aim of this article is to discuss the concept of mass basing on same fundamental principles of theoretical physics and considering general theory of relativity. Also to show that Eq. (2) is correct in some case and to show that the mass concepts, like the inertial and gravitational masses, are often misunderstood.

The oldest concept of mass is the one that was developed when people started trading and later that was used as matter content in body. With Newton's mechanics, the concepts of inertial and gravitational masses were beginning to use. Any physical quantity must be measurable in someway directly or indirectly. This perspective was concretized by Einstein when he constructed the Theory of Relativity. We can measure mass by using the laws of dynamics or with the scale. Therefore, matter content is nothing but the gravitational mass. In reality for the objects (not particles) a concept like matter content is not an exact thing and its measurement method is not different than that of gravitational mass. Therefore we would not deal with it in this article.

## 2. New concepts introduced by particle physics and special relativity: Energy, momentum and the problematic mass

The least action principle is one of the fundamental principles in physics and action give not absolute value of physics quantity, but show only how process proceed. But all physics processes in different inertial coordinate systems proceed similar. Therefore, in order to obtain the most general results, we take the condition that action stays constant under Lorentz transformations. By multiplying invariant quantity by a dimensional constant we can change its dimension to the dimension of action. We already have a quantity that is invariant under Lorentz transformations and a scalar, namely, the interval between events, $ds$ [3]. For the free particle in coordinate system where they located the interval is given as

$$ds = c\left(\sqrt{1-\frac{v^2}{c^2}}\right)dt. \tag{6}$$

Multiplying Eq. (6) by a dimensional constant (say $b$) and integrating we obtain the action as

$$S = -\int_{t_1}^{t_2} bc\left(\sqrt{1-\frac{v^2}{c^2}}\right)dt. \tag{7}$$

Here, the minus sign is there to guarantee that the extreme point of the integral is a minimum, instead of a maximum.

The Lagrangian is obtained by taking the time derivative of the action. In the Langrangian functional,

$$L = -bc\sqrt{1-\frac{v^2}{c^2}} \tag{8}$$

the fact that all the terms except the term $b$ are taken from the special relativity, makes the Langrangian valid for speeds close to the speed of light (and evidently for small velocities). Here, we are considering a particle that is not under any influence. Moreover, except its mass, such a particle hasn't got a physical property that is important for the problem under consideration. Therefore, $b$ should be related to mass (one can realize that since $b$ is constant, we have taken mass as a constant automatically). For a particle moving with small velocity, expanding the Langrangian in the ratio $v/c$ and discarding the small terms, we obtain

$$L = -bc\sqrt{1-\frac{v^2}{c^2}} \approx -bc + \frac{bv^2}{2c} \tag{9}$$

The constant terms of the Langrangian does not characterize the kinetics of the particle and can be omitted. Thus, to find the relation of the quantity $b$ with the mass of the particle we should look at the second term in the above expansion. The Langrangian for the free particle with small velocity is equal to its kinetic energy. Therefore,

$$\frac{bv^2}{2c} = \frac{mv^2}{2}. \tag{10}$$

This leads to $b = mc$. Then the relativistic Langrangian is

$$L = -mc^2\sqrt{1-\frac{v^2}{c^2}}. \tag{11}$$

Most generally, momentum is the partial derivative of the Langrangian with respect to its velocity

$$\mathbf{p} = \frac{\partial L}{\partial \mathbf{v}}, \qquad (12)$$

and to find the components of the momentum vector the partial derivative of Lagrangian with respect to each component of the velocity vector must be taken.

Now using Eq. (11) and Eq. (12), we can find the momentum of the free particle:

$$\mathbf{p} = \frac{m\mathbf{v}}{\sqrt{1 - \frac{v^2}{c^2}}}. \qquad (13)$$

Naturally, for small velocities, $\mathbf{v} \ll c$ this equation becomes the well known Newtonian formula

$$\mathbf{p} = m\mathbf{v}. \qquad (14)$$

In the above description that is based on special relativity the particle that is motionless or moving with small velocities has mass $m$ as seen from Eq. (14). There is no need to call mass $m_0$ in the nonrelativistic case because we treated mass as constant. With this approach the coefficient $1/(1 - v^2/c^2)^{1/2}$ has no physical relation with mass. In such approach the equation that is seen in many textbooks

$$m = \frac{m_0}{\sqrt{1 - \frac{v^2}{c^2}}} \qquad (15)$$

must be wrong because it is contain the term of rest mass. This mistake is come from some historical artefacts, which are outside the scope of particle physics and Einstein's theory of special relativity. The quantity mass in the Eqs. (11) and (13) is invariant under Lorentz transformations.

Eq. (13) is derived using the most important and trustworthy quantities and concepts of mechanics. This formula is valid for all speed and approaches infinity as the speed gets closer and closer to the speed of light. Why for 100 years (since the work of Lorentz in 1899) rest and relative mass concepts are often used? Is the mass of a particle, like electric charge, baryon and lepton numbers, an invariant quantity?

We can go back to the force concept that is known since Newton. Force is the derivative of the momentum with respect to time. The direction of the velocity of a particle, and the direction of the force on it can be quiet different. We can look at two different cases for demonstration: case 1: when the force and momentum are parallel (coincide) and case 2: when they are perpendicular. If the force and momentum of the particle are parallel, because of the force, the magnitude of the velocity changes, however the direction does not change. Then the derivative of the Eq. (13) with respect to time gives

$$\frac{d\mathbf{p}}{dt} = \frac{m}{\left(1-\frac{v^2}{c^2}\right)^{3/2}} \frac{d\mathbf{v}}{dt}. \tag{16}$$

If the force and momentum are perpendicular, the magnitude of the velocity does not change but the direction of the particle changes continuously. Then the derivative of the momentum becomes

$$\frac{d\mathbf{p}}{dt} = \frac{m}{\left(1-\frac{v^2}{c^2}\right)^{3/2}} \frac{d\mathbf{v}}{dt}. \tag{17}$$

Left hand side of this formula, which is described the change of momentum with respect to time (Eqs. (16) and (17)), are the forces and the right hand side is included acceleration of particle. In Newtonian physics, the ratios of the forces to the accelerations are masses. But, are these ratios

$$\frac{m}{\left(1-\frac{v^2}{c^2}\right)^{3/2}} \tag{18}$$

and

$$\frac{m}{\left(1-\frac{v^2}{c^2}\right)^{1/2}} \tag{19}$$

define the mass? No, Eqs. (18) and (19) are coefficients due to the relativistic motion that are not representing mass correctly. It is normal that for different situations coefficient in equations are different. However, it is not natural that the mass of the particle depends on the direction of the force acting on the particle. It is well known that these two definitions of mass were first introduced by Lorentz as longitudinal and transverse masses. Unlike Newtonian physics, here, mass is not the ratio of momentum and speed, because in relativistic and especially ultra-relativistic cases, inertia depends not only on the mass, but also on energy (for example kinetic energy). Without considering general relativity, we should not generalize the concept of mass in Newtonian physics.

Many authors use both the words "point particle" and "extended object" when dealing with similar problems in the frame of special relativity. In this section we only investigate the kinematics properties of free particles. In extended object particles are not free.

Recall that Eq. (11) describes the Lagrangian of a relativistic free particle. From this equation it is seen that as the velocity of the particle approaches the speed of light, the magnitude of L approaches zero and notice that L is always negative. Since the energy of a free particle is always larger than zero, in Eq. (11), the term, which describes the quantities related with the kinematics of the particle and which is positive should be increased by the addition of new term. This new term in the Lagrangian must also be valid in the Newtonian domain. The energy can be written as

$$E = \mathbf{p} \cdot \mathbf{v} - L. \tag{20}$$

If the insert the Lagrangian which is valid in Newtonian physics to Eq. (20), we obtain

$$E = \mathbf{p} \cdot \mathbf{v} - \frac{m v^2}{2} = \frac{m v^2}{2}. \tag{21}$$

This is the kinetic energy for the non-relativistic particle. The corresponding energy in relativistic case can be found by using the formulae (11) and (13)

$$E = \frac{mv^2}{\left(1 - \frac{v^2}{c^2}\right)^{1/2}} + mc^2 \left(1 - \frac{v^2}{c^2}\right)^{1/2} = \frac{mc^2}{\left(1 - \frac{v^2}{c^2}\right)^{1/2}}. \tag{22}$$

From this formula we can see that as the velocity of the free particle approaches the speed of light in vacuum, the energy becomes infinite. When the velocity of the free particle is zero, its energy is $mc^2$, not 0.

Now, to determine the relation between the energy and momentum of a particle in the frame of special relativity, we subtract the square of the Eq. (13) from the square of Eq. (22)

$$E^2 - p^2 c^2 = \frac{m^2 c^4 - m^2 v^2 c^2}{\left(1 - \frac{v^2}{c^2}\right)} = m^2 c^4 \tag{23a}$$

$$E^2 = p^2 c^2 + m^2 c^4, \tag{23b}$$

and

$$m^2 c^4 = E^2 - p^2 c^2. \tag{24}$$

From these formulas, it is seen again that the energy of a motionless particle (i.e. rest energy) is

$$E_0 = mc^2. \tag{25}$$

Using the same formulas (Eqs. (13) and (22)) we find

$$\mathbf{p} = \frac{E\mathbf{v}}{c^2}. \tag{26}$$

The particles having zero mass (photon, graviton), their speed in vacuum is $c$, so

$$\mathbf{p} = \frac{E}{c}. \tag{27}$$

Therefore, these objects even though they do not have mass, they have momentum and thus have pressure when the number of particles are high enough (because of the pressure is statistical quantity). Momentum, pressure and the inertia related not only to the particles (matter) with mass.

If we take $E/c$ as the fourth component of the 4-momentum, in the Lorentz transformations all the necessary conditions for the invariant 4-momentum would be satisfied. As known, components of the 4-momentum in two different coordinate systems can be written as

$$p_x = \frac{p'_x + \frac{v}{c^2}E'}{\sqrt{1 - \frac{v^2}{c^2}}}, \tag{28a}$$

$$p_y = p'_y, \tag{28b}$$

$$p_z = p'_z \tag{28c}$$

and

$$E = \frac{E' + v\, p'_x}{\sqrt{1 - \frac{v^2}{c^2}}}. \tag{29}$$

The Eq. (29) above is the simplified version of the fourth component of the 4-momentum

$$\frac{E}{c} = \frac{\frac{E'}{c} + \frac{v}{c} p'_x}{\sqrt{1 - \frac{v^2}{c^2}}}. \tag{30}$$

The invariant quantity in nature is the 4-dimensional momentum (or the energy-momentum tensor). Energy and momentum are not conserved separately. The conservation of momentum and conservation of energy separately is just an approximation and valid only for some of the cases within errors. We can compare Eq. (22) with the square of the interval between two events

$$ds^2 = c^2 dt^2 - dx^2 - dy^2 - dz^2 = c^2 dt^2 - dr^2.  \qquad (31)$$

Here, we see that also $ds^2$ is the square of a magnitude of a 4-vector, just like the $m^2 c^4$ term in Eq. (24). As $ds^2$ is invariant under Lorentz transformations, $m^2 c^4$ is also an invariant. That means, transforming from an inertial frame to any other one, the mass of the particle does not change. We see from this and the above that the mass of a particle is Lorentz invariant.

We can explain the problems in understanding of the concept of mass up to this point as historical artefacts. In the past, even some of the brilliant and famous physicists, were unable to understand which of the two different versions of the energy that introduced in 1900 by Poincare as Eq. (2) and in 1905 by Einstein as Eq. (1) is relevant. Moreover, the longitudinal and transverse masses that depend on the velocity of the particle (object) that Lorentz introduced in 1899 have been around. At those times longitudinal and transverse mass concepts were there. (It seems, even Einstein could only cope with these difficulties about 10 years after 1905. We should not forget that at those times there were no precise experiments on these subjects. General theory of relativity showed that attraction (gravitation) and inertia are properties of energy. Later, quantum physics (particle and nuclear physics) showed that the physical mass is a different and deeper concept. However, in some of the scientific journals and books the mass that depend on speed remained. This is mostly because this confusion about mass does not affect the results of the calculations.

**2. The mass of particles and objects defined like inertia and gravitation**

For the free particles that are considered as elementary until 1960's (we do not go down to the quark level here), the most important properties were their mass and rest energies ($E_0 = mc^2$). The velocity of the massless particles (photon and graviton) in vacuum is $c$. As the mass of elementary particles increase the number of properties that characterize them like electric charge, lepton number, baryon number, isospin, strangeness, etc. and the number of type of interactions they can have increase (A simple example for this is that neutrinos have only lepton numbers but electrons that are much heavier than neutrinos, have lepton numbers and also electric charges.). Now the question is this: Is the mass of these particles (or rest energies) conserved in any situations?

As known, the special theory of relativity describes the physical relation between space and time and determines the invariant physical quantities for the process involving any type of motion. This theory is one of the fundamental theories of physics that is valid where the gravitational field is constant and its gradient is zero, where the space can be considered as flat. Therefore, to understand the concept of mass, we should consider general relativity that has a more general scope.

**2.1. The effect of interaction fields on nonrelativistic particles**

Recall that in any interaction field (electrical, baryonic or gravitational) in which the particle is bound to the field ($E<0$), the mass of sub-atomic particles differs from their free states. Nuclear physics shows us that the free proton has a mass of 938.2 $MeV/c^2$ and four of them makes 3752.8 $MeV/c^2$. The mass of the free neutron is 939.6 $MeV/c^2$ when four nucleons come together to form the nucleus of the Helium atom consisting of two protons and two neutrons, about 28 MeV of energy is released. Therefore, the mass of the nucleons in the nucleus becomes less than the free ones. In general, in nuclear interactions, the mass of each nucleus cannot decrease more than $0.008\,mc^2$. Rainville et. al. [4] conducted a very precise and direct test of atomic mass conversion into photon energy and determined that the relation $E=mc^2$ is valid at least to a level of $\%0.00004$.

The binding energy of each baryon in a neutron star in its gravitational field is much higher than the binding energies in nuclear interactions. In some of the high mass and dense neutron stars the binding energies can be about $0.1\,mc^2$. In the gravitational field of black holes, the binding energy can exceed $0.2\,mc^2$. Evidently, for each sub-atomic particle the binding energy in Earth's, Sun's or normal star's gravitational fields is much less than the energy needed to form a nucleus.

Above discussion shows that the elementary particles loose some part of their mass in the attractive fields of the nucleus or stars. This shows that, in fact, mass is not a conserved quantity in general (unlike the charges of elementary particles). Contrary to the attractive forces, the repulsive forces between the baryons and the spin interactions of fermions increase the mass of the particles in the systems (Another example is when a photon tries to traverse a superconductor, it gains mass. Moreover, standard model of particle physics predicts that the fundamental particles obtain their mass through the interaction with the field).

Above, we have seen that the elementary particles loose more mass (rest energy) when becoming part of neutron stars and black holes. On the other hand, they gain considerable kinetic energy in rapidly rotating neutron stars (pulsars) or black holes. The newly born pulsar has a temperature about $10^{11} K$ and it can spin about hundreds of times per second. Because of these, for such a pulsar, the sum of thermal and rotational energy losses can be $(0.05-0.1) Mc^2$ (i.e. %5-10 of sum of all the rest energies of all the particles constituting the pulsar). Pulsars loose energy in general by emitting neutrino pairs under the collapse and neutron star formation, magnetodipole radiation, ultrarelativistic and relativistic particles.

Naturally, electron-positron pair annihilation which occurs due to electromagnetic (and sometimes due to weak) interaction has high probability of occurrence in the reverse direction as well (pair creation). These types of interactions show that in the interaction of elementary particles, even all the mass of particles can transform into radiation and radiation can transform completely into mass.

Therefore, the experiments on Earth and observations of the processes in the Universe prove that the formula $E = mc^2$ (for the bound particle) by Poincare in 1900 and the formula $E_0 = mc^2$ (for the free particle) by Einstein in 1905 are both correct. This means that these formulae explain the partial mass-loss (by electroweak and gravitational interactions) or complete mass-loss (by electroweak interaction) to radiation. Here, if we go beyond the concept of the mass that is related to the rest-energy of the particle, by the changing of mass with interaction, we can see that we have returned to the Poincare formula. Because, always in nuclear reactions the emitted energy is calculated using the formula $\Delta E = \Delta mc^2$.

The experimental and observational proofs of a formula do not necessarily show that all types of description of that formula or the ideas behind the formula are always correct and true. In the nuclear and electromagnetic interactions part of the mass of the particles transforms into radiation (until now we did not observe the gravitational binding energy transforming directly into radiation) and if the mass-loss is denoted by $\Delta m$, the emitted energy is $\Delta E = \Delta mc^2$. However, if the energy of the particle increases by $\Delta E$ its mass in general does not increase. For the bound particle $\Delta m = \Delta E/c^2$ formula is always correct and for radiation $m = h\nu/c^2$ formula is never correct.

Poincare formula ($E = mc^2$) might be partially correct since the masses of particle having high Fermi energy or thermal energy in compact star can not exceed its rest mass. This formula is valid only when the particle is bound since the kinetic energy of the bound particle does not increase the mass of the particle.

## 2.2. Definition of inertial and gravitational mass

In nuclear physics, high-energy physics and astrophysics, the type of matter and mass as used in chemistry always change. Sometimes matter changes into electromagnetic wave or the vice versa, it is created by electromagnetic wave. In all these processes the conserved quantity is neither the molecules or atoms themselves nor the elementary particles nor their masses. The conserved quantities are the different types of charge (electric charge, lepton number, baryon number).

In Einstein's theory of general relativity, the equivalence of inertial and gravitational masses is a postulate. Thus, there is only one mass in general relativity theory. Inertial and gravitational masses are different manifestations of the same mass.

At a point, or in a very small local region of space-time, general relativity gives the same results as special relativity. In such a case since there is the possibility to have no gravitational field, the need for the gravitational mass disappears for the particle (body) in the theory of special relativity. Therefore, in special relativity only one mass term, the inertial mass term can be used. In accelerators, magnetospheres of pulsars, shock waves of Supernova remnants and in other cosmic objects, the things that are accelerated to the velocity close to the speed of light are the elementary particles not even the atoms (except some ions of atoms). These particles have masses specific to themselves and these are the same masses that particle physicists use $\left(m = E_0/c^2\right)$ and this is the inertial mass of the particle. But, because of the mass does not change and in all cases the same, then there is no need to an "inertial mass" term. As the energy of the particle (body) increases its inertia increases. Inertia is a measure of energy. Since a particle (body) has a rest energy it has rest inertia too. This inertia, in Newtonian physics, is defined as if it is a property of the mass.

Cosmology shows that the radiation and any type of field gravitate. The gravitation is not only a property of mass, but it is mainly a property of energy. Therefore, if we take special and general theories of relativity and particle physics as fundamental, we can say that mass is not the only source of gravitational field, or the only source of inertia. Mass, having specific properties, is the quantity that corresponds to the minimum energy for the creation of free elementary particles (for example $\gamma + \gamma \rightarrow$ rest energy of any particle pair) or the quantity that corresponds to the rest energy ($m = E_0/c^2$). The energy of free particles can be changed continuously but their masses are discrete (for different types particles), well defined and specific to the particle. Only in this sense mass gains a different importance. When the energy of the particle (body) is close to its rest energy we are in the domain of Newtonian physics and we can consider the properties of energy as if they are properties of mass and this makes the calculations easier (we do not see $c^2$'s wandering around).

### 3. Conclusions

Only free elementary particles have constant rest energies and corresponding masses $(E_0/c^2 = m)$. The mass defined in this way has a special meaning in quantum (particle and nuclear) and in all physics. This mass is the ratio of the minimum value of energy needed for the creation of a particle to $c^2$. The value of minimum energy which is needed for the creation of a particle which is in a field and has negative energy is $E < E_0$. And in quantum states, this mass changes discretely. The particle in an atom that is isolated from external fields has a determined and constant mass.

It is more worth while and correct to consider the mass of the particle (body) as Lorentz invariant. In that case, the coefficient $(1 - v^2/c^2)^{1/2}$ for the particles moving with the relativistic speeds is not physically related to the mass of the particles as defined above. Then, in this case, same coefficient (with different exponents) that enters the equation for the relation between the forces and accelerations has no relation to mass.

When the energy of the particle is close to its rest energy, Newtonian physics become applicable and therefore we can consider the properties of the rest energy as if they are properties of mass and this makes the calculations easier by getting rid of $c^2$. For all objects in nature, when their heats and rotational or translational velocities change at the same time their energies, inertia and attraction change, whatever their states (gas, liquid or solid). However, these changes are much less than the energy related to the total mass of all the particles of the systems and therefore for the objects the Newtonian mass can be used. The gravitation properties and pressure of particles with the speeds of light depend on their energies. All types of energy (rotation, heat and the energy of fields) gravitate and in the most general sense have inertial properties.